\documentclass[12pt]{article}
\usepackage{amstex}
\usepackage{amssymb}

\newfont{\bit}{cmbxti10 scaled 1728}

\renewcommand{\thefootnote}{\fnsymbol{footnote}}

\begin{document}
\newpage
\pagestyle{empty}

\begin{center}
{\LARGE {Colombeau's Generalized Functions\\
 on \\
Arbitrary Manifolds\\ 
}}

\vspace{3cm}

{\large Herbert BALASIN
\footnote[2]{e-mail: hbalasin @@ phys.ualberta.ca}
\footnote[3]{supported by the APART-program of the 
Austrian Academy of Sciences}
}\\ 
{\em
 Institute for Theoretical Physics, University of Alberta\\
 Edmonton, T6G 2J1, CANADA
}\\[.8cm]
\end{center}
\vspace{2cm}
\begin{abstract}
We extend the Colombeau algebra of generalized functions
to arbitrary (paracompact) $C^\infty$ $n$-manifolds $M$. 
Embedding of continuous functions and distributions is achieved 
with the help of a family of $n$-forms defined on the tangent bundle $TM$,
which form a partition of unity upon integration over the fibres.

\end{abstract}

\rightline{Alberta-Thy-35-96}
\rightline{TUW 96 -- 20}
\rightline{August 1996}

\renewcommand{\thefootnote}{\arabic{footnote}}
\setcounter{footnote}{0}
\newpage  
\pagebreak
\pagenumbering{arabic}
\pagestyle{plain}
\renewcommand{\baselinestretch}{1}
\small\normalsize 
\section*{\Large\bit Introduction}
Colombeau theory \cite{Col1,ArBi} set out to give a mathematically consistent
way of multiplying distributions. 
From a physical point of view the Colombeau algebra provides a sound framework
to accommodate calculations involving regularization methods employed
to deal with singular quantities that actually arise as products
of distributions. The most prominent example in this regard is provided 
by the renormalization procedure of perturbative quantum field theory.
Recently, there have been some attempts \cite{Par,BaNa1,BaNa2,Ba1,LoSo,ClVi}
to apply a similar formalism to spacetime singularities that arise in general
relativity. Unless one is willing to restrict to manifolds
that are topologically ${\mathbb R}^n$ the application of the
Colombeau formalism requires its generalization to arbitrary
manifolds.

Although it is a pretty recent development the definition of the 
Colombeau algebra has undergone various changes, starting from a functional
analytic motivation, involving the delicate subject of differential
calculus in locally convex vector spaces \cite{Col0}.

The definition \cite{Col2} we are going to generalize in this work considers 
the elements of the Colombeau algebra ${\mathcal G}({\mathbb R}^n)$
as moderate one-parameter families $(f_\epsilon)$ of $C^\infty$ functions 
denoted by $C_M^\infty({\mathbb R}^n)$ up to negligible families denoted by 
$C_N^\infty({\mathbb R}^n)$, where the adjectives
refer to certain growth conditions in the parameter $\epsilon$ of the family.
The embedding of continuous functions and distributions
into algebra is achieved with the help of a smoothing-kernel, which has 
to obey certain properties.
The above quotient is then such that 
$C^\infty({\mathbb R}^n)$ becomes a subalgebra of 
${\mathcal G}({\mathbb R}^n)$ thereby reconciling the two different 
embeddings as constant sequences or with the aid of the smoothing kernel.

Although the definition of the Colombeau algebra used above
is fairly straightforwardly generalized to an arbitrary manifold \cite{Ba1},
the notion of the smoothing kernel presents some difficulties
since it draws heavily upon concepts specific to ${\mathbb R}^n$.
On the other hand the smoothing kernels play an important role
for the embedding of distributions as linear subspaces into the Colombeau
algebra. Although there have been proposals \cite{CoMe} which 
weaken the condition on the moments, we will stick to the original
condition by taking the smoothing kernel to be an $n$-form on the 
tangent-bundle $TM$ of the manifold $M$ (more precisely a family
of $n$-forms of this type, which becomes a partition of unity upon
integration along the fibre).
Since diffeomorphisms of $M$ act linearly on the fibres of $TM$ 
this approach gives (invariant) meaning to the conditions on the moments 
familiar from ${\mathbb R}^n$. \
The embedding of smooth and continuous functions with the aid of
the above smoothing kernel produces locally defined moderate 
functions, whose sum defines the corresponding Colombeau-object.

Our work will be organized as follows. In chapter one
we recall the basic notions of the Colombeau algebra.
Section two will be devoted to a brief survey of distribution theory
on arbitrary manifolds, thereby giving a manifestly covariant
formulation of the latter. Finally in chapter three we present
our generally covariant formulation of the Colombeau algebra 
motivated by the extension of smoothing kernels which we will
consider to be a suitable family of $n$-forms on the $2n$-dimensional 
tangent-bundle.

\section*{\Large\bit 1) Moderate and negligible functions, association
and all that} 

The basic idea of Colombeau's approach for the multiplication
of distributions is to find a (differential) algebra large
enough to contain all the usual $C^\infty$ functions as a subalgebra 
and the distributions as a linear subspace. The construction
starts by considering one-parameter families $(f_\epsilon)$ of 
$C^\infty$ functions subject to the condition

\begin{eqnarray}\label{Mod}
  && C^\infty_M = \{ (f_\epsilon)\vert f_\epsilon \in 
    C^\infty({\mathbb R}^n)\quad \forall K \subset {\mathbb R}^n compact, 
    \forall \alpha\in {\mathbb N}^n\quad  \\
  &&\hspace*{1cm}\exists\, N\in {\mathbb N},\exists\> \eta > 0,\exists\> c>0 
    \quad s.t. \sup_{x\in K}\vert D^\alpha f_\epsilon(x)\vert \leq 
  \frac{c}{\epsilon^N}\quad\forall 0<\epsilon< \eta\},\nonumber\\
&&\mbox{where}\quad D^\alpha = 
\frac{\partial^{|\alpha|}}{(\partial x^1)^{\alpha_1}
\cdots(\partial x^n)^{\alpha_n}}.\nonumber
\end{eqnarray}
One way of getting some intuition for these objects is by
considering them as regularizations of  functions that
(possibly) become singular in the limit $\epsilon\to 0$. All operations like
addition and multiplication are defined pointwise, and one easily
proves that (\ref{Mod}) is an algebra, which will be denoted
$C^\infty_M$. 

$C^\infty$-functions are canonically embedded into $C_M^\infty$
as constant sequences, whereas continuous functions (and distributions)
of at most polynomial growth
require a smoothing kernel $\varphi \in {\mathcal S}$ 
\footnote{$\mathcal S$ 
denotes the space of rapidly decreasing $C^\infty$ functions}.
Since it represents an approximate $\delta$-function one requires that
\begin{equation}\label{smooth}
  \int d^nx \varphi(x)=1 \qquad \int d^nx\> x^\alpha \varphi(x)=0 \quad
  |\alpha|\geq 1
\end{equation}
The embedding is done by convoluting the (rescaled and shifted)
smoothing kernel $\varphi$ with $f$, i.~e.~
$$
f_\epsilon(x) = \int d^ny \frac{1}{\epsilon^n} \varphi(\frac{y-x}{\epsilon})
f(y).
$$

In order to reconcile the different embeddings of $C^\infty$ functions
one identifies them by employing a suitable ideal 
$C^\infty_N({\mathbb R}^n)$. Its members will be addressed as negligible
functions in the following.
\begin{eqnarray}\label{Neg}
  && C^\infty_N({\mathbb R}^n) = \{ (f_\epsilon)\vert (f_\epsilon) \in 
  C^\infty_M({\mathbb R}^n)\quad
  \forall K \subset {\mathbb R}^n compact, \forall \alpha\in {\mathbb N}^n,  
   \forall N\in {\mathbb N}\nonumber\\
  &&\hspace*{1cm}\exists\> \eta > 0,\exists\> c>0,\quad 
  s.t. \sup_{x\in K}\vert D^\alpha f_\epsilon(x)\vert \leq 
  c\epsilon^N\quad\forall\> 0<\epsilon< \eta\}
\end{eqnarray}
Considering the difference $f(x) - \int d^ny (1/\epsilon^n) 
\varphi(\frac{y-x}{\epsilon})f(y)$, where $f$ denotes an
arbitrary $C^\infty$ function of at most polynomial growth, one easily 
checks that it is a negligible function.
The Colombeau algebra ${\mathcal G}({\mathbb R}^n)$ is therefore defined to be
the quotient of  $C^\infty_M({\mathbb R}^n)$ with respect to
$C^\infty_N({\mathbb R}^n)$. A Colombeau object is thus a moderate family
$(f_\epsilon(x))$ of $C^\infty$ functions modulo negligible families.

The usual distribution theory arises from coarse graining the Colombeau
algebra employing an equivalence relation called association. 
Two Colombeau objects $(f_\epsilon)$ and $(g_\epsilon)$ 
will be considered associated if
\begin{equation}\label{assoc}
\lim_{\epsilon\to 0} \int d^nx (f_\epsilon(x) -g_\epsilon(x))\varphi(x) =0
\qquad \forall \varphi \in {\mathcal D}.
\end{equation}
This equivalence relation is much coarser than the one used for the definition
of $\mathcal G$. It is compatible with addition, differentiation and 
multiplication by $C^\infty$ functions. It is, however not compatible with
multiplication. Intuitively speaking different Colombeau objects are 
packaged together into one association-class. One might think of such a class
as containing different regularizations of the same (possibly singular) 
non-smooth function. 

Let us give a simple (by now classical \cite{Col2}) example showing the 
power of the association calculus. Consider $\theta^n$ which as 
piecewise continuous function gives rise to the same (regular) distribution 
as $\theta$. Upon naive differentiation one obtains
$$
\theta^n(x) = \theta(x) \Rightarrow n \theta(x) \theta'(x) = \theta'(x) 
\Rightarrow \theta(0) = \frac{1}{n},
$$
which is a contradiction since $\theta(0)$ is independent of $n$.
With regard to the Colombeau algebra $\theta^n$ is  no longer equal to 
$\theta$, they are however associated. Since association respects
differentiation we have
$$
\theta^n(x) \approx \theta(x) \Rightarrow n \theta^{n-1}(x) \theta'(x) 
\approx \theta'(x) \Rightarrow \theta^n(x)\theta'(x) \approx 
\frac{1}{n+1}\delta(x),
$$
which only tells us that we are allowed to replace $\theta^n\theta'$ by
$\delta/(n+1)$. Since multiplication breaks association we do not
encounter any ambiguities.

\section*{\Large\bit 2) Distributions on arbitrary manifolds}

The main goal of this section is to give a manifestly covariant formulation
of distribution theory, which shows that the distribution concept
relies only on the differentiable structure of the underlying manifold $M$
and does not require any additional notions such as the existence of a metric
or a volume-form \cite{ChBr,BaNa1}.
Usually distributions on ${\mathbb R}^n$ are defined as elements of the 
(topological) dual of test-function space $\mathcal D$, which in the simplest 
case is  considered to consist of all $C^\infty$ functions with compact 
support. Locally integrable functions $f(x)$ are embedded into distribution 
space as so-called regular functionals
\begin{equation}\label{regular}
(f,\varphi) = \int d^n x f(x)\varphi(x).
\end{equation}
Operations like differentiation and multiplication by arbitrary
$C^\infty$ functions are defined via the corresponding operations
on $\mathcal D$, namely
\begin{eqnarray*}
(D^\alpha f,\varphi) &:=& (-)^{|\alpha|}(f,D^\alpha \varphi)\quad \alpha \in 
{\mathbb N}^n \\
( g\> f,\varphi) &:=& (f,g \> \varphi),
\end{eqnarray*} 
which reduce to standard integral relations if one restricts to regular
distributions generated by differentiable functions. Both of the above
operations are well-defined since they map $\mathcal D$ onto itself.

In order to generalize the above concepts to arbitrary manifolds $M$, we first
have to decide what to do about test-function space. Although $C^\infty$
functions with compact support are a concept that makes sense on arbitrary
manifolds this would not allow us to embed locally  integrable functions
in the same way as we did in ${\mathbb R}^n$, where we made use of the
natural volume form $d^n x$, unless we are willing to single out a volume
form. Thinking, however, of $\varphi$ and $d^nx$ as parts of {\em one} 
object, we are immediately lead to consider $C^\infty$ $n$-forms 
$\tilde{\varphi}$ with compact support as the natural generalization of  
test-functions. Taking into account that the latter are sections of a 
vector-bundle and by employing a partition of unity  (which requires 
the underlying manifold $M$ to be paracompact) we basically may construct 
the locally convex vector-space topology in very much the same way as in 
${\mathbb R}^n$. Distributions are then once again defined as the elements 
of the topological dual of this space.

The concepts of multiplication by $C^\infty(M)$ functions is without problems.
In order to generalize differentiation we make use of the Lie-derivative
along an arbitrary $C^\infty $ vector-field and the classical 
identity 

$$
(L_X f,\tilde{\varphi}) = \int_M L_X f \tilde{\varphi} =
\int_M d (i_X (f \tilde{\varphi})) - \int_M f L_X\tilde{\varphi} = 
-(f,L_X\tilde{\varphi})
$$
where the last equality is taken to be true for arbitrary distributions.
Let us once again emphasize that the embedding of locally integrable
functions led us to generalizing test-function space to $C^\infty$
$n$-forms with compact support. 

\section*{\Large\bit 3) Generally covariant formulation of the \\Colombeau 
algebra}  

This chapter is devoted to a formulation of the Colombeau algebra
suitable for arbitrary manifolds $M$. Considering the definition of
moderate and negligible functions (\ref{Mod},\ref{Neg}) one sees that they
immediately generalize to $M$, since the conditions required for the 
definition remain invariant under coordinate transformations.
A manifestly covariant definition may be given with the aid
of the Lie-derivative 
\begin{eqnarray*}
C^\infty_M(M) &=& \{ (f_\epsilon)| f_\epsilon \in C^\infty(M)  \forall\>
K \subset M\> compact, 
\forall\>\{X_1,\dots,X_p\}\>\\
&&\hspace*{0.5cm}X_i\in\Gamma(TM),[X_i,X_j ]=0,\exists N\in{\mathbb N},\exists 
\eta> 0, 
\exists c>0\> \\
&&\hspace*{0.5cm} s.t. \sup_{x\in K}| L_{X_1}\dots L_{X_p}
f_\epsilon(x)|
\leq \frac{c}{\epsilon^N}\quad 0<\epsilon<\eta\},
\end{eqnarray*}

\begin{eqnarray*}
C^\infty_N(M) &=& \{ (f_\epsilon) \in C^\infty_M(M) | \forall\>
K \subset M\> compact,\forall\>\{X_1,\dots,X_p\}\> \\
&&\hspace*{0.5cm}X_i\in\Gamma(TM),
 [X_i,X_j]=0,\forall q\in{\mathbb N},\exists \eta> 0, \exists c>0\>\\
&&\hspace*{0.5cm} s.t. \sup_{x\in K}| L_{X_1}\dots
L_{X_p}f_\epsilon(x)|
\leq c\epsilon^q\quad 0<\epsilon<\eta\}.
\end{eqnarray*}

However, looking at the smoothing kernel $\varphi$ required for the 
embedding of continuous functions (and distributions) one realizes that the 
condition on the moments does not remain invariant under coordinate 
transformations. One way of remedying this situation was presented in 
\cite{CoMe}, where instead of a fixed smoothing kernel $\varphi$ a whole 
family $\varphi_\epsilon$ was considered, which allowed a weakening of 
the conditions on the moments.

We will try to follow a different path, which allows us to keep the
conditions on the moments in a coordinate invariant manner. 
This seemingly paradoxical statement is easily understood in terms of 
the tangent bundle $TM$ of $M$. Let us consider a bundle-atlas of $TM$ 
induced by an atlas $M$ and let $(x,\xi)$ denote the respective coordinates 
of an arbitrary chart. Now fix a differential $n$-form 
\begin{equation}\label{smoothfrm}
\tilde{\varphi} = \varphi(x,\xi)(d\xi^1+ N^1{}_idx^i)\wedge\cdots\wedge
(d\xi^n+ N^n{}_idx^i)
\end{equation}
which we require to obey
\begin{eqnarray}\label{smoothprp}
  &&\int\limits_{T_xM} i_x^*\tilde{\varphi} = \int \varphi(x,\xi) d^n\xi 
  =1\qquad\int\limits_{T_xM} \xi^\alpha i_x^* \tilde{\varphi} = 
  \int \xi^\alpha\varphi(x,\xi) d^n\xi = 0\\
  &&\mbox{where }\> i_x:T_xM\to TM\qquad
  i_x^*\tilde{\varphi}=\varphi(x,\xi)d^n\xi\nonumber\\
  &&\hspace*{2cm}\xi\mapsto (x,\xi)\nonumber
\end{eqnarray}
The advantage of the tangent-bundle formulation comes from
the fact that diffeomorphisms in $M$ induce a specific type
of diffeomorphism on $TM$, namely fibre-preserving bundle morphisms,
which act {\em linearly} on the fibres thereby leaving (\ref{smoothprp})
invariant. Moreover, the rescaling and shift operations are now naturally 
interpreted as action of the structure group on $TM$, given by
\begin{eqnarray}\label{structgr}
&&\phi_\epsilon:TM\to TM\qquad (x,\xi)\mapsto (x,\frac{1}{\epsilon}\xi),
\nonumber\\
&&\phi_a:TM\to TM\qquad (x,\xi)\mapsto (x,\xi+a)
\end{eqnarray}
which are specific $IGL(n,{\mathbb R})$\footnote{we actually
consider the fibres $T_xM$ to be affine spaces} transformations.
Let us now use the smoothing-form $\tilde\varphi$ to embed
a continuous function $f$
$$
f_\epsilon(x):= \int\limits_{T_xM}\phi_\epsilon^*\phi_{-x}^*i_x^*
\tilde{\varphi}f
= \int d^n\xi\frac{1}{\epsilon^n} \varphi(x,\frac{\xi-x}{\epsilon})f(\xi)=
\int d^n\xi \varphi(x,\xi)f(x+\epsilon\xi).
$$
The last relation makes explicit use of the coordinate representation
of the function $f$ with respect to a given coordinate chart
(which we assume to be ${\mathbb R}^n$) in order to lift $f$
to a function on $T_xM$. This entails that $f_\epsilon$ is
also only defined locally. However, using a $C^\infty$ partition
of unity $(\rho_i)$ subordinate to the cover $(U_i)$ allows us to patch 
the local objects together to a global one
\begin{equation}\label{glemb}
  f_\epsilon(x) := \sum\limits_i \rho_i(x)f_{i,\epsilon}(x),
\end{equation}
where the sum is always finite due to the local finiteness of the
cover $(U_i)$. Note that actually every term in the sum is
a well-defined $C^\infty$--object on $M$.
The above construction may be completely absorbed into the
$n$-form $\tilde{\varphi}$ by taking a family
of $n$-forms $\tilde{\varphi}_i$ subordinate to the cover $(U_i)$ 
such that 
$$
\sum\limits_i \int\limits_{T_x M}i_x^*\tilde{\varphi}_i = 1
$$
That is to say $\rho_i(x) := \int\limits_{T_x M}i_x^*\tilde{\varphi}_i$
defines a partition of unity subordinate to the (locally finite) 
cover $(U_i)$. 
It is now easy to show that the two different embeddings
of $C^\infty$ functions differ only by elements belonging to $C^\infty_N(M)$.
$$
f_\epsilon(x)-f(x)=\sum\limits_i \int d^n\xi\varphi_i(x,\xi)
(f(x+\epsilon \xi) - f(x)) = {\mathcal O}(\epsilon^q),
$$
where the last equality is derived from Taylor-expanding $f(x+\epsilon \xi)$.
Although the embedding (\ref{glemb}) depends on the atlas employed,
since we made specific use of the local representation of the
function in order to lift it to the tangent-bundle, this dependence 
disappears for $C^\infty$ functions
\begin{eqnarray*}
  \bar{f}_\epsilon(\bar{x})&=&\int d^n\bar{\xi} 
  \bar{\varphi}(\bar{x},\bar{\xi})\bar{f}(\bar{x}+\epsilon\bar{\xi})
  =\int d^n\xi\varphi(x,\xi)f(\mu^{-1}(\mu(x)+\epsilon
  \frac{\partial\mu}{\partial x}\xi))\\
  &=&\int d^n\xi\varphi(x,\xi)f(x+\epsilon \xi) + {\mathcal O}(\epsilon^q)
  =f_\epsilon(x) + {\mathcal O}(\epsilon^q),
\end{eqnarray*}
where once again the result was obtained by Taylor-expanding up to
order $q$.
\vfill

\noindent
{\bf Acknowledgement:} The author wants to thank Michael Oberguggenberger
and Michael Kunzinger for introducing him to the subject of Colombeau theory,
during his stay in Innsbruck. Furthermore the author wants to thank
the relativity and cosmology group at the University of Alberta and especially
Werner Israel, for their kind hospitality, during the final stage
of this work.
\newpage

\section*{\large\bit Conclusion}
In this paper we generalized the Colombeau algebra to arbitrary 
manifolds. The main problem that had to be overcome was the 
embedding of continuous functions and distributions with the
help of a smoothing kernel. The definition of the latter used in
the standard ${\mathbb R}^n$-approach required all its moments to vanish.
Unfortunately, this concept does not remain invariant under coordinate
transformations and does therefore not generalize to arbitrary
manifolds. Taking advantage of the tangent bundle we were, 
however, able to maintain the condition on the moments by
taking the smoothing kernel to be a differential $n$-form
defined on the $2n$-dimensional tangent bundle. The coordinate 
invariance of this construction is guaranteed by linear action
of $M$-diffeomorphisms on the fibres of $TM$.

\end{document}